\documentclass[conference]{IEEEtran}
\IEEEoverridecommandlockouts
\usepackage{amsmath,amssymb,amsfonts}
\usepackage{cite}
\usepackage{graphicx}
\usepackage{psfrag}
\usepackage{subfigure}
\usepackage{amsmath}
\usepackage{array}
\usepackage{algorithmicx}
\usepackage{algpseudocode}
\usepackage{setspace}
\usepackage{blindtext}
\usepackage{url}
\usepackage{epstopdf}
\usepackage{verbatim}
\usepackage{color}
\usepackage{amsmath}
\usepackage{bm}
\usepackage{multirow}
\usepackage[table,xcdraw]{xcolor}
\usepackage{booktabs}
\usepackage{makecell}
\usepackage{ntheorem}
\usepackage{textcomp}
\usepackage{amsmath}

\newtheorem{Prob}{Problem}

\newcommand{\RNum}[1]{\uppercase\expandafter{\romannumeral #1\relax}}

\theoremseparator{:}

\title{A Novel GCN based Indoor Localization System with Multiple Access Points}
\author{Yanzan Sun, Qinggang Xie, Guangjin Pan, \IEEEauthorblockN{Shunqing Zhang\thanks{\IEEEauthorrefmark{1} Shunqing Zhang is corresponding author.}\IEEEauthorrefmark{1}, and Shugong Xu}\\
Shanghai Institute for Advanced Communication and Data Science, \\
Key laboratory of Specialty Fiber Optics and Optical Access Networks, \\
Joint International Research Laboratory of Specialty Fiber Optics and Advanced Communication, \\
Shanghai University, Shanghai, 200444, China\\

Email: \{yanzansun, victoriaxg, guangjin\_pan, shunqing, shugong\}@shu.edu.cn}

\begin{document}
\maketitle

\begin{abstract}
With the rapid development of indoor location-based services (LBSs), the demand for accurate localization keeps growing as well. To meet this demand, we propose an indoor localization algorithm based on graph convolutional network (GCN). We first model access points (APs) and the relationships between them as a graph, and utilize received signal strength indication (RSSI) to make up fingerprints. Then the graph and the fingerprint will be put into GCN for feature extraction, and get classification by multilayer perceptron (MLP).In the end, experiments are performed under a 2D scenario and 3D scenario with floor prediction. In the 2D scenario, the mean distance error of GCN-based method is 11m, which improves by 7m and 13m compare with DNN-based and CNN-based schemes respectively. In the 3D scenario, the accuracy of predicting buildings and floors are up to 99.73\% and 93.43\% respectively. Moreover, in the case of predicting floors and buildings correctly, the mean distance error is 13m, which outperforms DNN-based and CNN-based schemes, whose mean distance errors are 34m and 26m respectively.
\end{abstract}

\begin{IEEEkeywords}
localization, graph convolutional network, received signal strength indication, multi-layer perceptron
\end{IEEEkeywords}

\IEEEpeerreviewmaketitle
\section{Introduction} \label{sect:intro}
The rapid growth of smart city \cite{smartcity1,smartcity2} and smart mobile terminals have triggered high-precision indoor localization requirements. Although satellite positioning systems, such as global positioning system (GPS) \cite{GPS}, or Beidou navigation system \cite{Jiang2018Beidou}, have been developed with sub-meter level outdoor localization accuracy, they can hardly achieve the same level in the indoor environment due to satellite signal occlusions. To address this issue, the indoor localization technologies utilize more diversified wireless signals, including wireless fidelity (WiFi)\cite{wifi1}, Bluetooth low energy (BLE)\cite{ble}, and increasingly popular 3GPP LTE/5G technology \cite{LTE}. Among the existing indoor localization schemes, there are in general two mainstream algorithms, namely {\em the geometric based} and {\em the fingerprint based}. The geometric based approaches measure the propagation delays and directions of wireless signals, and calculate the corresponding locations using geometric theorems. Typical algorithms include direct localization or three-point localization methods \cite{disbased} using time of arrival (TOA), time difference of arrival (TDOA), time of flight (TOF), angle of arrival (AOA) \cite{AOA,RBF2010,TOA}. The fingerprint based approaches, however, establish the fingerprint database during the offline training stage and estimate the target localization in the online inference stage based on real time measurements \cite{Guo2020fingerprint,Wang2017fingerprint,Karabey2015fingerprint}, where received signal strength indication (RSSI) and channel state information (CSI) are commonly adopted as the fingerprint indicator. 
 
Although the fingerprint based schemes provide a more accurate localization results in general \cite{xiang2019}, the bottleneck is to find a nonlinear relationship between the target location and the corresponding fingerprints. Conventional machine learning algorithms, such as K-nearest neighbor (KNN) \cite{KNN}, weighted K-nearest neighbor (WKNN) \cite{WKNN} and restricted Boltzmann machine (RBM) \cite{RBM}, suffer from high computational complexity in the online stage, which are rarely used in the practical deployment. Recently, with the development of deep learning, convolutional neural network (CNN) \cite{chen2017confi}, deep residual sharing learning \cite{resloc}, and recurrent neural networks (RNNs) \cite{RNN} have been proposed to learn this nonlinear relation and achieved sub-meter level localization accuracy while maintaining a reasonable implementation complexity simultaneously. 

The recent proposed graph convolutional network (GCN) \cite{GCN} provides another dimension to extract the features from non-Euclidean data structures. Since the fingerprint based localization system generates the non-Euclidean data structures with different propagation environments, a straightforward question is whether we can directly apply GCN to exploit the geometric relations of different access points (APs) and extract the intrinsic features inside. Motivated by this idea, we propose an indoor localization scheme with high accuracy for multiple APs by fully utilizing the feature extraction capabilities of GCN architecture. As demonstrated through some numerical results, the proposed scheme can achieve 11m mean distance error in the 2D scenario. Besides, it can be up to 99.73\% and 93.43\% accuracy of predicting buildings and floors respectively in the 3D scenario, and achieve 13m mean distance error at the same time. Both experiment results outperform the Baseline localization algorithms. The main contributions of our work are listed below.

\begin{itemize}
\item{\em Hybrid GCN and MLP Structure} To fully exploit the geometric relations among multiple APs, we consider multiple APs and the collected RSSIs as an undirected graph, and extract the intrinsic features using GCN. By combining the extracted geometric features among multiple APs and the original RSSI observations, we develop a MLP structure to understand the mapping relation inside. Compared with other deep learning approaches, our proposed scheme achieves better localization accuracy as shown later.  
\item{\em Adjacency Matrix Construction} Since the adjacency matrix is the key design parameter for GCN, we propose two different methodologies to construct the adjacency matrix to represent geometric relations among different APs. With the AP deployment information, we propose to use the Euclidean distances among APs to construct the adjacency matrix. When the deployment information is unavailable, we propose to use the statistical information of user received signals from multiple APs instead, which achieves promising localization accuracy as well. The adjacency matrix construction methods in this paper are not optimal, but will be further studied in our future work.
\end{itemize}

The remainder of this paper is organized as follows. In Section~\ref{sect:sys}, we introduce some preliminary information, and the GCN based problem formulation is discussed in Section~\ref{sect:prob}. The detailed network configuration of GCN-MLP structure, as well as the adjacency matrix construction, are provided in Section~\ref{sect:nndesign}. In Section~\ref{sect:experiment}, we present our experimental results and the concluding remarks are given in Section~\ref{sect:conc}.

\section{Preliminaries} \label{sect:sys}
In this section, we briefly introduce the system model and the localization procedures, followed by some preliminary knowledge about GCN.

\begin{figure}
\centering
\includegraphics[width = 3.4 in]{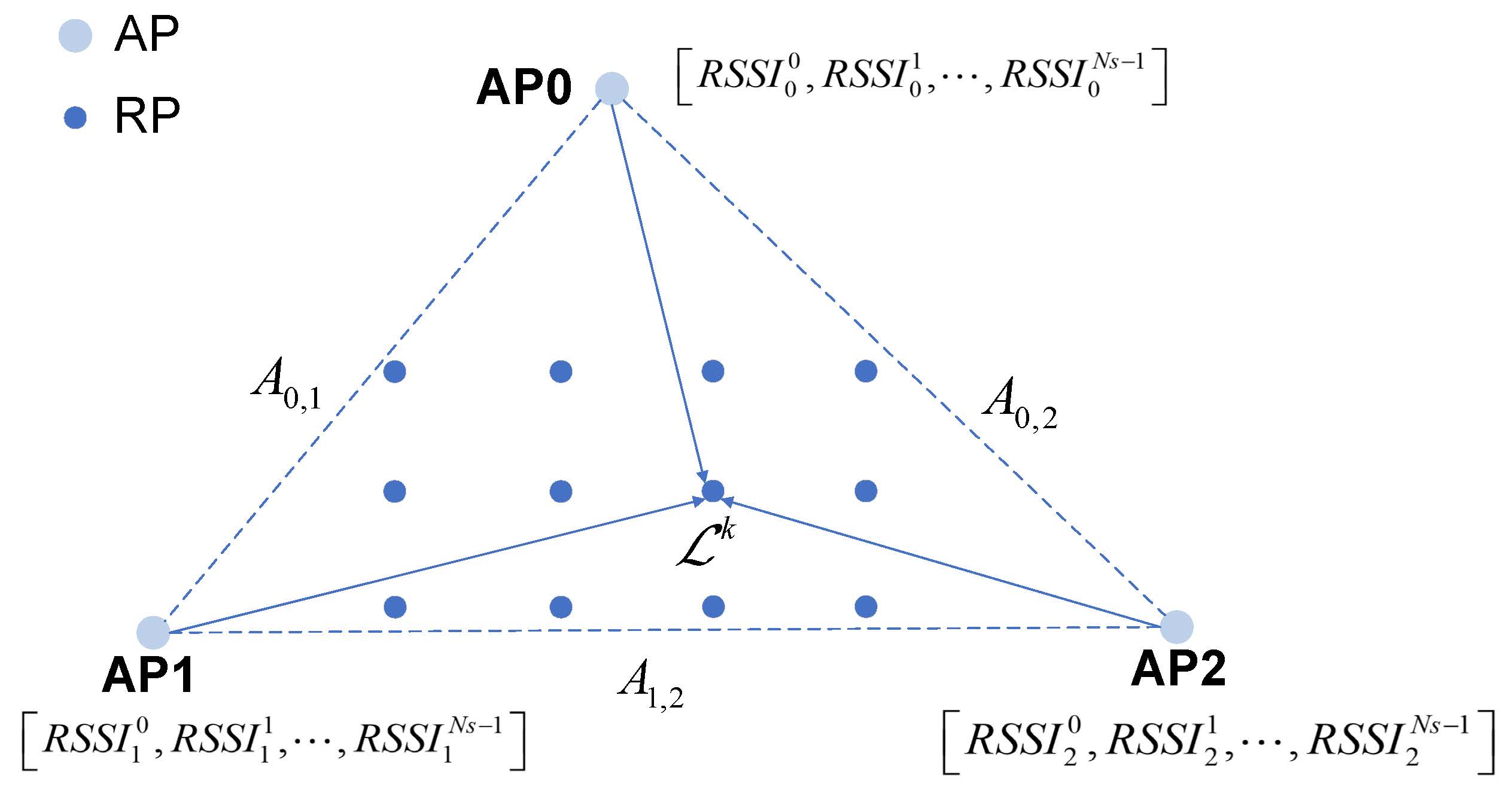}
\caption{This is the way RPs and APs generate fingerprints. $RSSI^t_s$ denotes the sampled RSSI at the $t^{th}$ time slot from the $s^{th}$ AP. And $A_{i,j}$ represents edges of two APs. }
\label{fig:fingerprint}
\end{figure}

\subsection{System Model and Localization Procedures}
\label{subsect:prob}
Consider an indoor localization system with $N_{AP}$ APs and $N_{RP}$ reference points (RPs) as shown in Fig.~\ref{fig:fingerprint}
. Denote $\mathcal{L}^{RP} = [\mathcal{L}^{1}, \ldots, \mathcal{L}^{N_{RP}}]^{T}$ to be the locations of $N_{RP}$ RPs, and the location $\mathcal{L}$ in terms of $\mathcal{L}^{RP}$ is given by $\mathcal{L} = \mathbf{p}(\mathcal{L}) \cdot \mathcal{L}^{RP}$, where $\mathbf{p}(\mathcal{L}) = [p^1(\mathcal{L}), \ldots, p^{N_{RP}}(\mathcal{L})]$ can be regarded as the likelihood probability. 

In the offline training stage, we first construct an RSSI fingerprint database $\mathcal{DB} = \left\{(\mathbf{X}(\mathcal{L}^{k}), \mathbf{1}^{k})\right\}$, where $\mathbf{X}(\mathcal{L}^{k})$ is the sampled RSSIs of consecutive $N_s$ time slots at the location $\mathcal{L}^{k}$ and $\mathcal{L}^{k} = \mathbf{1}^{k} \cdot \mathcal{L}^{RP}$ since $\mathbf{1}^{k}$ is the unit vector with the $k^{th}$ element equals to $1$ and $0$ otherwise. $\mathbf{X}(\mathcal{L}^{k})$ contains the sampled RSSIs of consecutive $N_s$ time slots from $N_{AP}$ APs and the mathematical representation is given by,
\begin{small}
\begin{eqnarray}
\ \mathbf{X}(\mathcal{L}^{k}) =\left[\begin {array}{ccc} 
{RSSI}_{0}^{0}(\mathcal{L}^{k}) &  \cdots&  {RSSI}_{0}^{N_s-1}(\mathcal{L}^{k})\\
{RSSI}_{1}^{0}(\mathcal{L}^{k}) &  \cdots&  {RSSI}_{1}^{N_s-1}(\mathcal{L}^{k})\\
\vdots & \vdots & \vdots\\
{RSSI}_{N_{AP}-1}^{0}(\mathcal{L}^{k}) & \cdots&  {RSSI}^{N_s-1}_{N_{AP}-1}(\mathcal{L}^{k})\\
\end{array}
\right]
\end{eqnarray}
\end{small}
where ${RSSI}_{s}^{t}(\mathcal{L}^{k})$ denotes the sampled RSSI at the $t^{th}$ time slot from the $s^{th}$ AP. 

In the online deployment stage, we estimate the location $\hat{\mathcal{L}}$ based on the aforementioned fingerprint database $\mathcal{DB}$ and the real time observation $\mathbf{X}(\mathcal{L})$, e.g., 
\begin{eqnarray}
\hat{\mathcal{L}} = f\left(\mathbf{X}(\mathcal{L}),\mathcal{DB}
\right).
\end{eqnarray}
where the function $f(\cdot)$ denotes the mapping relation from the joint space of fingerprint database $\mathcal{DB}$ and real time observation $\mathbf{X}(\mathcal{L})$ to the estimated location $\hat{\mathcal{L}}$.

\subsection{Preliminary Knowledge about GCN} \label{sect:GCN}

\begin{figure}
\centering
\includegraphics[width = 3.4 in,height=1.6 in]{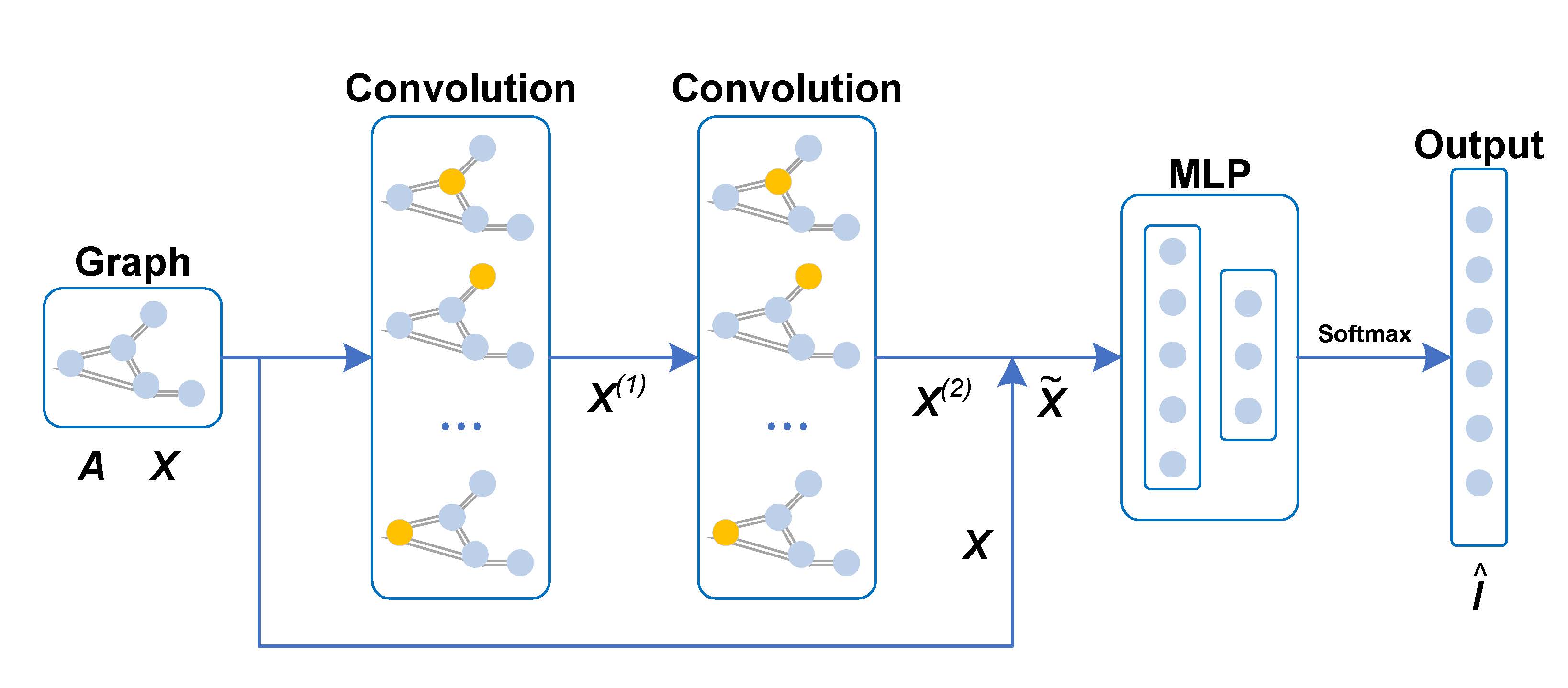}
\caption{The architecture of GCN, which includes two convolution layers and MLP with three fully connected (FC) layers. Outputs, $\mathbf{X}^{(1)}$ and $\mathbf{X}^{(2)}$, of two convolutional layers and original graph input MLP together to be classified.}
\label{fig:architecture}
\end{figure}

The topological structure of APs in the practical environment is in general non-Euclidean. Traditional neural networks, including CNN and RNN, often fail to characterize this non-Euclidean relation, since graph can be irregular, a graph may have different number of nodes, edges and neighbors, resulting in some operations (e.g., convolution) are hard to compute in graph domain \cite{wu2020comprehensive}. In this paper, we use GCN instead and jointly consider the wireless propagation properties in the indoor environment, and the basic concepts of GCN are elaborated as follows.

A undirected graph $\mathcal{G}$ can be described by a collection of nodes $\mathcal{V} = \{v_1, \ldots, v_{N}\}$ and edges $\mathcal{A} = \{A_{i,j}\}_{\forall i,j \in \mathcal{V}}$. Due to the undirected nature of graph $\mathcal{G}$,  $\mathcal{A}$ is symmetrical. Denote $\mathcal{D}$ to be the diagonal degree matrix with $\mathcal{D} = \textrm{diag}\left[\{\sum_{j}({A}_{i,j})\}, \forall i\right]$, and the normalized graph Laplacian matrix $\mathbf{L}_{\mathcal{G}}$ is equal to $\mathbf{I}_{N} - \mathcal{D}^{-\frac{1}{2}}\mathcal{A}\mathcal{D}^{-\frac{1}{2}}$, where $\mathbf{I}_{N}$ is the identity matrix with dimension $N$. Since $\mathbf{L}_{\mathcal{G}}$ is positive semi-definite, it can be factored as $\mathbf{L}_\mathcal{G} = \mathbf{U}\mathbf{\Lambda}\mathbf{U}^T$, where $\mathbf{\Lambda} = \textrm{diag} [\lambda_1, \cdots, \lambda_N]$ is a diagonal matrix with eigenvalues. According to the convolution theorem \cite{GCN}, the graph convolution can be written as,
\begin{eqnarray}
\label{eqn:conv_theorem1}
\bm{g} *_{(G)} \bm{X}(\mathcal{L}) = \mathbf{U}\big(\mathbf{U}^T\bm{g} \odot \mathbf{U}^T\bm{X}(\mathcal{L})\big),
\end{eqnarray}
where $\bm{g}$ is graph filter, $*_{(G)}$ is graph convolution, and $\odot$ denotes dot product. If we denote a filter as $\bm{g}_\beta(\mathbf{\Lambda})=\mathbf{U}^T\bm{g}$, \eqref{eqn:conv_theorem1} can be simplified as
\begin{eqnarray}
\label{eqn:conv_theorem2}
\bm{g}_\beta *_{(G)} \bm{X}(\mathcal{L}) = \mathbf{U}\bm{g}_\beta\mathbf{U}^T\bm{X}(\mathcal{L}) = \mathbf{U}\bm{g}_{{\beta}}(\mathbf{\Lambda})\mathbf{U}^T\bm{X}(\mathcal{L}).
\end{eqnarray}

Since the graph convolution is usually complexity prohibited, the graph filter operation is approximated by Chebyshev polynomials through \cite{Defferrard2016Convolutional},
\begin{eqnarray}
\bm{g}_{{\beta}}(\mathbf{\Lambda} ) \approx \sum_{i=0}^{I-1}{\mathbf{{\theta}}_i}\mathbf{T}_{i}(\tilde{\mathbf{\Lambda }}),
\end{eqnarray}
where $\tilde{\mathbf{\Lambda}} = 2 \mathbf{\Lambda} / \lambda_{\max} - \mathbf{I}_{N}$, $\lambda_{\max} = \max \{\lambda_i,i=1,\cdots ,N\}$. $\mathbf{T}_i(x) = 2x\mathbf{T}_{i-1}(x)-\mathbf{T}_{i-2}(x)$ denotes Chebyshev polynomials, with the initial conditions given by $\mathbf{T}_{0}(x)=1$ and $\mathbf{T}_{1}(x)=x$.
$\{\mathbf{\theta}_i\}$ denote the corresponding coefficients vector. As $\mathbf{T}_i(\tilde{\mathbf{L}}_\mathcal{G}) =\mathbf{U} \mathbf{T}_i(\tilde{\mathbf{\Lambda}})\mathbf{U}^T$, \eqref{eqn:conv_theorem2} can be well approximated by,
\begin{eqnarray}
\bm{g}_{{\beta}} *_{(G)} \bm{X}(\mathcal{L})\approx \sum_{i=0}^{I-1}{\mathbf{{\theta}}_i}\mathbf{T}_{i}(\tilde{\mathbf{L}}_\mathcal{G})\bm{X}(\mathcal{L}).
\end{eqnarray}
In the above expression, $\tilde{\mathbf{L}}_\mathcal{G} = 2 \mathbf{L}_\mathcal{G} / \lambda_{\max} - \mathbf{I}_{N}$. If $I = 2$ and $\lambda_{\max} = 2$, the convolution equation can be simplified as,
\begin{eqnarray}
\bm{g}_{{\beta}} *_{(G)} \bm{X}(\mathcal{L}) \approx  \mathbf{\theta}\big(\mathbf{I}_N + \mathcal{D}^{-\frac{1}{2}}\mathcal{A}\mathcal{D}^{-\frac{1}{2}}\big)\bm{X}(\mathcal{L}).
\end{eqnarray}

Therefore, one layer of the GCN can be represented as, 
\begin{eqnarray}
\mathbf{X}^{(l+1)} 
=g_l(\mathbf{X},\mathcal{A})= \sigma(\tilde{\mathcal{A}}\mathbf{X}^{(l)}\mathbf{\Theta}^{(l)}),
\label{eqn:H_l+1}
\end{eqnarray}
where $\tilde{\mathcal{A}} = \mathbf{I}_N + \mathcal{D}^{-\frac{1}{2}}\mathcal{A}\mathcal{D}^{-\frac{1}{2}}$, $\sigma(\cdot)$ is the nonlinear activation function, and $\mathbf{X}^{(l)}$ is feature of the $l^{th}$ layer with $\mathbf{X}^{(0)} = \bm{X}(\mathcal{L})$. In summary, GCN can study graph features and relationships between nodes with help of Laplacian eigenvalues and eigenvectors.

\section{GCN based Problem Formulation}
\label{sect:prob}

The GCN based scheme, as illustrated before, can be utilized to extract the in-depth features across the RSSI observations from different APs. If we regard APs to be the graph nodes $\mathcal{V}_{AP}$ and the relationships between neighboring APs to be edges $\mathcal{A}_{AP}$, we can construct a GCN based feature representation $\widetilde{\mathbf{X}}(\mathcal{L})$ as provided in Section~\ref{sect:GCN}, and obtain,
\begin{eqnarray}
\widetilde{\mathbf{X}}(\mathcal{L}) = \mathbf{X}(\mathcal{L})- \mathbf{X}^{(2)}(\mathcal{L}), \label{eqn:fea}
\end{eqnarray}
where the GCN filtered features $\mathbf{X}^{(1)}(\mathcal{L})$ and $\mathbf{X}^{(2)}(\mathcal{L})$ are defined as,
\begin{eqnarray}
\mathbf{X}^{(1)}(\mathcal{L}) & = & g_{1}\left(\mathbf{X}(\mathcal{L}),\mathcal{A}_{AP}\right), \label{eqn:fea1} \\
\mathbf{X}^{(2)}(\mathcal{L}) & = & g_{2}\left(\mathbf{X}^{(1)}(\mathcal{L}),\mathcal{A}_{AP}\right).\label{eqn:fea2}
\end{eqnarray}

With the above GCN based manipulations, we can approximate the original localization function $\hat{\mathcal{L}} = f\left(\mathbf{X}(\mathcal{L}),\mathcal{DB}
\right)$ as,
\begin{eqnarray}
\hat{\mathcal{L}} = g\left(\widetilde{\mathbf{X}} (\mathcal{L})\right), \label{eqn:est}
\end{eqnarray}
where $\widetilde{\mathbf{X}} (\mathcal{L})$ is defined by \eqref{eqn:fea}, \eqref{eqn:fea1} and \eqref{eqn:fea2}. By introducing the subscript $m$, we can define the mean absolute error (MAE) minimization problem as follows.

\begin{Prob}[\em MAE Minimization] The overall MAE minimization problem of the proposed GCN based localization scheme is given by,
\begin{eqnarray}
\underset{g(\cdot), g_1(\cdot), g_2(\cdot)}{\textrm{minimize}} && \frac{1}{M}\sum_{m=1}^{M} \|\hat{\mathcal{L}}_{m} - \mathcal{L}_{m}\|_2
\label{eqn:mini1}\\
\textrm{subject to} && \hat{\mathcal{L}}_m = g\left(\widetilde{\mathbf{X}} (\mathcal{L}_m)\right), \nonumber \\
&& \widetilde{\mathbf{X}}(\mathcal{L}_m) = \mathbf{X}(\mathcal{L}_m)- \mathbf{X}^{(2)}(\mathcal{L}_m), \nonumber \\
&& \mathbf{X}^{(1)}(\mathcal{L}_m) = g_{1}\left(\mathbf{X}(\mathcal{L}_m),\mathcal{A}_{AP}\right), \nonumber \\
&& \mathbf{X}^{(2)}(\mathcal{L}_m) = g_{2}\left(\mathbf{X}^{(1)}(\mathcal{L}_m),\mathcal{A}_{AP}\right).
\end{eqnarray}
where $\|\cdot\|_2$ denotes the $l_2$ norm of the inner vector.
\end{Prob}

The optimal solution of the above minimization problem is in general difficult to figure out, since the exhaustive search over all the possible linear or nonlinear functions $g(\cdot)$ are required. 
Therefore, we transform the original MAE minimization problem to a standard classification problem, where it computes the likelihood distribution with respect to several RP, which reduce the searching complexity. In order to measure the classification performance, we utilize cross-entropy between $\mathbf{p}(\mathcal{L}_m)$ and $\mathbf{p}(\hat{\mathcal{L}}_m)$, which is illustrated mathematically as follow.

\begin{Prob}[\em Cross-Entropy Minimization] The cross-entropy minimization problem of the proposed GCN based localization scheme is given by,
\begin{eqnarray}
\underset{g_{\textrm{MLP}}(\cdot), \theta^{(0)}, \theta^{(1)}}{\textrm{minimize}} && - \frac{1}{M}\sum_{m=1}^{M} \mathbf{p}(\mathcal{L}_m)^{T} \log \left(\mathbf{p}(\hat{\mathcal{L}}_m)\right)
\label{eqn:mini2}\\
\textrm{subject to} && \mathbf{p}(\hat{\mathcal{L}}_m) = g_{\textrm{MLP}}\left(\widetilde{\mathbf{X}} (\mathcal{L}_m)\right), \nonumber \\
&& \widetilde{\mathbf{X}}(\mathcal{L}_m) = \mathbf{X}(\mathcal{L}_m)- \mathbf{X}^{(2)}(\mathcal{L}_m), \nonumber \\
&& \mathbf{X}^{(1)}(\mathcal{L}_m) = \sigma \left( \tilde{\mathcal{A}}_{AP} \mathbf{X}(\mathcal{L}_m) \mathbf{\Theta^{(0)}} \right), \nonumber \\
&& \mathbf{X}^{(2)}(\mathcal{L}_m) = \sigma \left(\tilde{\mathcal{A}}_{AP}\mathbf{X}^{(1)}(\mathcal{L}_m) \mathbf{\Theta^{(1)}}\right).
\nonumber\end{eqnarray}
where $g_{\textrm{MLP}}(\cdot)$ defines the MLP processing for classification as proposed in \cite{xiang2019}.
\end{Prob}

\section{GCN based Localization Scheme}
\label{sect:nndesign}
In this section, we propose the GCN based localization scheme and the corresponding neural network architecture to solve the above cross-entropy minimization problem. Rather than consider all APs to be the same in the existing technologies, we tend to make full use of the topological structure of APs and make GCN learn features from it. In order to further describe the relationships among APs, we provide a comparative study on the adjacency matrix $\mathcal{A}_{AP}$ design.

\subsection{Neural Network Configuration}
\label{subsect:NNarchitecture}
To improve the ability of extracting features, we design the neural network with two convolutional layers GCN and three FC layers MLP. The detail of configuration is shown as Table~\ref{tab:parameter}. Firstly, we put the fingerprint database and adjacency matrix into GCN, which is the key to extract intrinsic features. The input size of the first convolutional layer of GCN is $N \times n_s$, which means RP receives signals from $N$ APs for $n_s$ time slots. Besides, the input size of the next layer is equal to the output size of the prior layer. Then, the features from GCN will get classified by MLP. And the input of MLP consists of two, the features extracted by GCN and the original graph data, as shown in Fig.\ref{fig:architecture}, so that the input size of MLP is $N\times n_s$. After feature extraction of GCN and classification of MLP, the output size is $1\times K$, which represents the probabilities of corresponding TAs or RPs.

In addition, we choose ReLU as the activation function of GCN and MLP except for the last FC layer of MLP, whose activation function is Softmax to output the results of classification. ReLU and Softmax are shown as below,
\begin{equation}
    ReLU(z_i) = max(0,z_i),
\end{equation}
\begin{equation}
    Softmax(z_i) = \frac{e^{z_i}}{\sum_{j=0}^{j=K-1}e^{z_j}},
\end{equation}
where $z_i$ is the $i^{th}$ element of output, $\mathbf{X}^{(l)}$ of $l^{th}$ layer.
\begin{table} [ht]
\centering
\caption{An Overview Of Network Configuration and Parameters.}
\label{tab:parameter}
\footnotesize
\renewcommand\arraystretch{1.5}
	\begin{tabular}{|c|c|c|c|c|}  
		\hline
        \bottomrule
		Functions & Layers & Input Shape & Output Shape & Activation \\ 
		\hline \bottomrule
    	\multirow{2}{0.8cm}{\centering{GCN}}
		& GCN 1 & $N\times n_s$ & $ N\times n_s $ & ReLU\\         \cline{2-5}
		& GCN 2 & $N\times n_s $ & $N\times n_s$ & ReLU\\     \hline
		\bottomrule
		\multirow{6}{0.8cm}{\centering MLP}
		& Flatting & $N\times
		 n_s$ & $N n_s $ & \\
		\cline{2-5}
		& FC 1 & $N n_s$ & $ 64 $ &   ReLU \\
		\cline{2-5}
		&FC 2&  $ 64 $& 32& ReLU\\  \cline{2-5}
		&FC 3 & 32 & K& Softmax\\  \hline \bottomrule		
	\end{tabular}
\end{table}

\subsection{Adjacency Matrix Selection}
\label{subsect:adjacent}
Considering the adjacency matrix $\mathcal{A}_{AP}$ design directly determines the performance of the GCN algorithm, we propose two adjacency matrix construction methods according to the intuitive physical relationship among APs.

For the scenario of a large positioning area with many APs, most APs have no direct relationship and the corresponding adjacency matrix is sparse. It is difficult to construct the adjacency matrix especially when the APs positions are unknown. In such a scenario, we propose to construct the adjacency matrix based on the statistical profile of the training dataset and calculate the probability of receiving the $i^{th}$ and $j^{th}$ AP signal simultaneously from the training dataset. Therefore, the element of the adjacency matrix $\mathcal{A}_{AP}$ can be expressed as, 
\begin{equation}
A_{i,j}=A_{j,i}=\frac{N_{i,j}}{M_{train}}, 
\end{equation}
where $M_{train}$ is the total number of training data sets, and ${N_{i,j}}$ is the number of data sets that can receive the $i^{th}$ and $j^{th}$ AP signal at the same time.

For the scenario of a small positioning area, the user equipment can receive signals from all APs. We define the adjacency matrix element as the the reciprocal of distance between the $i^{th}$ and $j^{th}$ AP signals. Therefore, the element of the adjacency matrix $\mathcal{A}_{AP}$ can be expressed as, 
\begin{equation}
A_{i,j}=A_{j,i}=\frac{1}{d_{ij}}
\end{equation}

By choosing different values of the adjacency matrix, the proposed GCN based positioning system can adapt to various environments and be widely used. These two adjacency matrix construction methods above are not necessarily optimal. Other construction methods suitable for different scenarios will be further studied and analyzed in our future works.

\section{Experiment Results} \label{sect:experiment}
In this section, we test our proposed GCN-based fingerprint positioning algorithm on the public database UJIIndoorLoc  \cite{Torres2014UJIIndoorLoc} in the 2D and 3D localization scenarios. The multi-building and multi-floor WiFi data from UJIIndoorLoc is utilized as the training and testing data, where 520 different APs are spread over three buildings with 4 or 5 floors. A total of 21,048 sampled RSSI values are captured in 993 reference points, of which 19,937 samples are used for training and 1,111 samples for testing. If the AP signal is available, the RSRP values are recorded as the fingerprints, otherwise, the fingerprint is marked as -110dB in our experiments. Therefore, we set $n_s=1$, $N_{AP}=520$, and use the first probability based adjacency matrix construction method to obtain the adjacency matrix $\mathcal{A}_{AP}$ in this case. The following numerical results are provided to show the effectiveness of our proposed algorithm. 

\subsection{Localization Accuracy}  
\begin{figure}
\centering
\includegraphics[width = 3 in,height=2.3in]{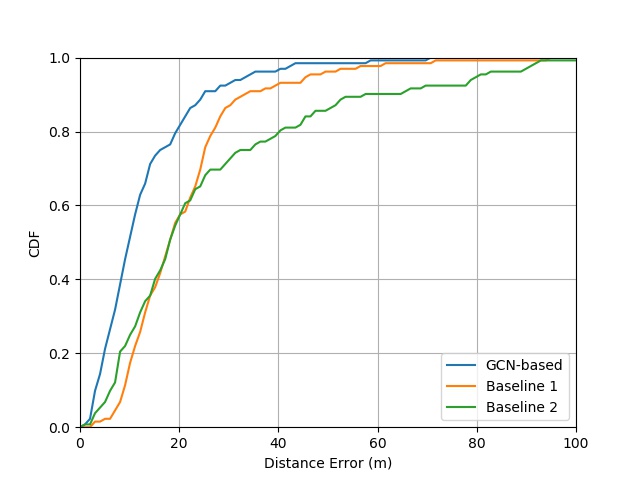}
\caption{CDF of prediction distance error in 2D scenario.}
\label{fig:CDF2D}
\end{figure}

We compare the proposed GCN-based method with two classic network based systems, e.g., Baseline 1: DNN-based method with four FC layers and Baseline 2: CNN-based method with three convolutional layers and two pooling layers. To evaluate the localization performance of GCN-based method in the 2D scenario, we pick the single floor data from the UJIIndoor database to train three different neural networks for 1500 epochs, and then predict the estimated coordinates as network output. The cumulative distribution function (CDF) of distance errors is shown as Fig.~\ref{fig:CDF2D}. We can find that the 60\% distance error of the proposed GCN-based method is about 10m, which outperforms those of Baseline 1 and Baseline 2. Additionally, the mean distance errors of GCN-based method, Baseline 1 and Baseline 2 are 11m, 18m and 24m, respectively. From the above results, we can conclude that the proposed method significantly improves localization accuracy compared with Baseline 1 and Baseline 2, considering that the chosen adjacency matrix can provide prior information about relationships among APs.

\subsection{Extension to 3D Scenarios} 
In this part, we extend the experiment to 3D scenarios, and the entire UJIIndoorLoc database is utilized to test the ability of the GCN-based method as well. In this experiment, the expected output of the proposed algorithm consists of the predicted building, floor and coordinates. The accuracy of buildings and floors prediction is shown in Table~\ref{tab:PreAccuracy}. Although all three neural networks can predict the floors and buildings results with high accuracy, GCN-based method achieves slightly better performance than Baseline 1 and Baseline 2.

\begin{figure}
\centering
\includegraphics[width = 3in, height=2.3in]{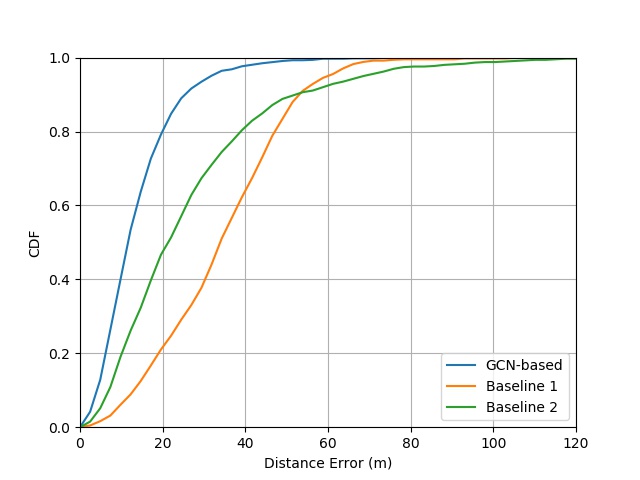}
\caption{CDF of prediction distance error in 3D scenario.}
\label{fig:CDF3D}
\end{figure}

\begin{table} [ht]
\centering
\caption{Accuracy of predicting floors and buildings.}
\label{tab:PreAccuracy}
\footnotesize
\renewcommand\arraystretch{1.5}
	\begin{tabular}{|c|c|c|}  
		\hline  
        \bottomrule
		Networks&Floor&Building \\ 
		\hline \bottomrule
		GCN-based& 0.9343 &0.9973\\ 
		Baseline 1& 0.9235 &0.9973\\ 
		Baseline 2& 0.9271 &0.9955\\     \hline
		\bottomrule
	\end{tabular}
\end{table}
In the case of predicting floors and buildings correctly, we calculate the distance error of predicted coordinates of three methods. To evaluate the performance of the proposed fingerprint localization algorithm, the CDF of localization error is illustrated in Fig.~\ref{fig:CDF3D}. We find that the 60\% distance error of the proposed GCN-based method is about 12m, which performs better than Baseline 1 and Baseline 2 with the 60\% distance error of 38m and 25m respectively. In addition, the mean distance errors of GCN-based method, Baseline 1 and Baseline 2 are 13m, 34m and 26m respectively, which shows significant improvement in terms of localization accuracy.

\begin{figure}[h] 
	\centering  
	\vspace{-0.35cm} 
	\subfigtopskip=2pt 
	\subfigbottomskip=2pt 
	\subfigcapskip=-5pt 
	\subfigure[Statistics of 2D distance error]{
		\label{fig:Boxplot2D}
		\includegraphics[width=0.45\linewidth]{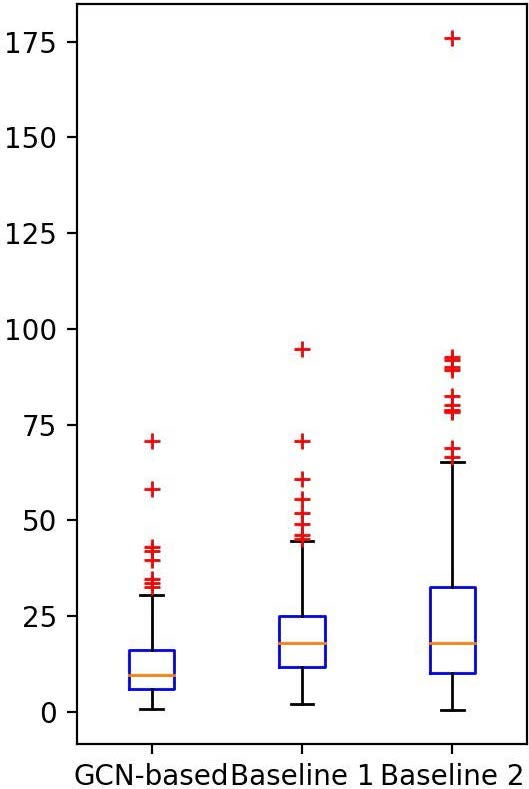}}
	\quad 
	\subfigure[Statistics of 3D distance error]{
		\label{fig:Boxplot3D}
		\includegraphics[width=0.45\linewidth]{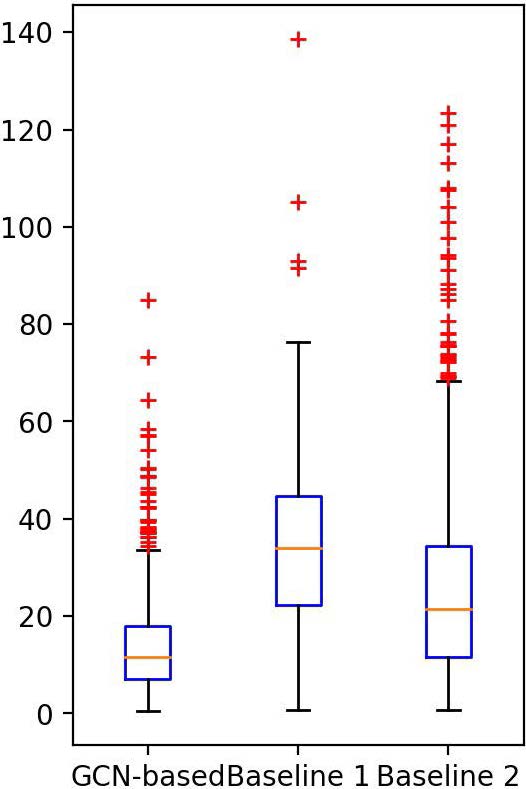}}
	\quad
	\caption{Statistics of distance error in 2D and 3D scenario.}
	\label{fig:Boxplot}
\end{figure}

To better show the profiles of positioning results, we summarize the above experiment results in form of boxplot diagrams in Fig.~\ref{fig:Boxplot}, which show minimum, maximum and median values, 25-th and 75-th percentiles, and possible outliers. It’s obvious that distance errors of GCN-based method decrease significantly compared with Baseline 1 and Baseline 2 in both 2D and 3D scenarios. And GCN-based method has a more concentrated distribution, which means the proposed method can provide a more reliable positioning service in both 2D and 3D scenarios with the aid of the adjacency matrix.

\section{Conclusion} \label{sect:conc}
In this paper, we propose a new indoor localization algorithm based on GCN. Different from traditional localization algorithms based on neural network, we first extract features from graph, which consists of APs and the relationship among them. Then the fingerprint is established with RSSIs. After that, the outputs of GCN together with the original graph data are put into MLP to get the classified results. Finally, we verify the indoor localization algorithm based on GCN shows higher precision compared to DNN and CNN respectively.

\section*{Acknowledgement}
This work was supported by the National Natural Science Foundation of China (NSFC) under Grants 62071284, 61871262, 61901251 and 61904101, the National Key Research and Development Program of China under Grants 2019YFE0196600, the Innovation Program of Shanghai Municipal Science and Technology Commission under Grant 20JC1416400, Pudong New Area Science \& Technology Development Fund, and research funds from Shanghai Institute for Advanced Communication and Data Science (SICS)..

\bibliographystyle{IEEEtran}
\bibliography{IEEEabrv,bb_rf}

\end{document}